
\input harvmac

\def\dag{\dagger}
\def\VEV{VEV}
\noblackbox
\baselineskip=16pt

\Title{\vbox{\baselineskip12pt{\hbox{CTP-TAMU-1/93}}%
{\hbox{hep-ph/9302220}}}}
{\vbox{\centerline{Vortex Solutions in Two-Higgs Systems}
\medskip
\centerline{and $\tan\beta$}}}
\centerline{HoSeong ~La\footnote{$^*$}{%
e-mail address: hsla@phys.tamu.edu, hsla@tamphys.bitnet}   }

\bigskip\centerline{Center for Theoretical Physics}
\centerline{Texas A\&M University}
\centerline{College Station, TX 77843-4242, USA}
\vskip 0.7in

Classical vortex solutions in various two-Higgs systems are studied.
The systems we consider include the standard model with two Higgs doublets,
in which case the vortex appears as part of a string-like object.
The Higgs potentials contain several different couplings in general and
the spontaneous symmetry breaking involves with two
different vacuum expectation values.
In particular it is shown that the existence of such a solution in general
requires a specific ratio of the two Higgs vacuum expectation values,
i.e. $\tan\beta$, and some inequalities between different Higgs couplings.
This ratio can be determined in terms of the couplings in the Higgs potential.
The Higgs masses are also computed in this case.
(1+2)-d solutions are topological so that they are topologically stable and the
Bogomol'nyi bound is saturated for some couplings.
Some comments on the stabilization of (1+3)-d solutions are also given.
Thus, as long as such a defect can be formed in the early universe, stable or
not, $\tan\beta$ is no longer an independent free parameter in the theory.

\baselineskip=16pt

\Date{1/93} 
 \noblackbox

\def\hat{\widehat}
\def\tilde{\widetilde}

\def\la{\lambda}
\def\gt{{\tilde g}}
\def\half{{\textstyle{1\over 2}}}

\def\e{{\rm e}}
\def\pa{\partial}
\def\mbox#1#2{\vcenter{\hrule \hbox{\vrule height#2in
		\kern#1in \vrule} \hrule}}  
\def\lfr#1#2{{\textstyle{#1\over #2}}}
\def\eps{\epsilon}

\def\rbr{\right)}
\font\cmss=cmss10 \font\cmsss=cmss10 scaled 833
\def\IZ{\relax\ifmmode\mathchoice
{\hbox{\cmss Z\kern-.4em Z}}{\hbox{\cmss Z\kern-.4em Z}}
{\lower.9pt\hbox{\cmsss Z\kern-.4em Z}}
{\lower1.2pt\hbox{\cmsss Z\kern-.4em Z}}\else{\cmss Z\kern-.4em Z}\fi}
\def\IR{\relax\ifmmode\mathchoice
{\hbox{\cmss I\kern-.5em I}}{\hbox{\cmss R\kern-.5em R}}
{\lower.9pt\hbox{\cmsss I\kern-.5em I}}
{\lower1.2pt\hbox{\cmsss R\kern-.5em R}}\else{\cmss R\kern-.5em R}\fi}
\def\tr{{\rm tr}}

\def\cos{{\rm cos}}
\def\sin{{\rm sin}}

\def\CL{{\cal L}}

\vfill\eject

\newsec{Introduction}

The idea of the spontaneous symmetry breaking is one of the most important
building blocks of the electroweak theory, which is one of the greatest
achievements in theoretical particle physics.
Although there is no doubt about the validity of the electroweak theory,
the final experimental proof however is not yet completed because
the wanted Higgs particle is still at large. And also there is some possibility
that the detail of such spontaneous symmetry breaking may be a bit different
from the minimal content originally proposed.
There is growing anticipation, although indirect, the electroweak symmetry
breaking may be induced by two Higgs doublets, rather than one.

For example,
the recent measurements of the gauge couplings\ref\PDG{Particle Data Group,
Phys. Rev. {\bf D45} (1992).}\ have led us to anticipation
that the minimal supersymmetric Grand Unified Theories (GUTs)\ref\rSGUT{S.
Dimopoulos and H. Georgi, Nucl. Phys. {\bf B193} (1981) 150; N. Sakai, Zeit. f.
Phys. {\bf C11} (1981) 153.}\
or supergravity GUTs\ref\rSGGT{A.H. Chamseddine, R. Arnowitt and P. Nath, Phys.
Rev. Lett. {\bf 49} (1982) 970; L.E. Ibanez, Phys. Lett. {\bf 118B} (1982) 73;
J. Ellis, D.V. Nanopoulos and Tamvakis, Phys. Lett. {\bf 121B} (1983) 123;
K. Inoue, A. Kakuto, H. Komatsu and S. Takeshita, Prog. Theo. Phys. {\bf 68}
(1982) 927; L. Alvarez-Gaum\'e, J. Polchinski and M.B. Wise, Nucl. Phys. {\bf
B221} (1983) 495; J. Ellis, J.S. Hagelin, D.V. Nanopoulos and Tamvakis, Phys.
Lett. {\bf 125B} (1983) 275; L. Iba\~nez and C. Lopez, Nucl. Phys. {\bf B233}
(1984) 545; L.E. Iba\~nez, C. Lopez and C. Mu\~nos, Nucl. Phys. {\bf B250}
(1985) 218.}\ref\revSGUT{For reviews see P. Nath, R. Arnowitt and A.H.
Chamseddine, ``{\it Applied N=1 Supergravity}," (World Sci., Singapore,
1984); H.P. Nilles, Phys. Rep. {\bf 110} (1984) 1; H. Haber and G. Kane, Phys.
Rep. {\bf 117} (1985) 75.}\
with the supersymmetry scale of order 1 TeV or below may be a
phenomenologically plausible unified theory of strong and electroweak
interactions\ref\rUNCoup{P. Langacker, in {\it Proc. PASCOS 90 Symposium} ed.
by P. Nath and S. Reucroft (World Sci., Singapore, 1990); P. Langacker and M.
Luo, Phys. Rev. {\bf D44} (1991) 817; J. Ellis, S. Kelley and D.V. Nanopoulos,
Phys. Lett. {\bf 249B} (1990) 441; {\bf 260B} (1991) 131; U. Amaldi, W. de Boer
and H. F\"ursteanu, Phys. Lett. {\bf 260B} (1991) 447.}.
These supersymmetric GUTs in general require at least two Higgs
multiplets for the electroweak symmetry breaking\revSGUT%
\ref\rtwoH{N.G. Deshpande and E. Ma, Phys. Rev. {\bf D18} (1978) 2574;
 J.F. Donoghue and L.-F. Li, Phys. Rev.
{\bf D19} (1979) 945; H.E. Haber, G.L.Kane and T. Sterling, Nucl. Phys. {\bf
B161} (1979) 493; E. Golowich and T.C. Yang, Phys. Lett. {\bf 80B} (1979) 245.
}\ref\GerH{H. Georgi, Hadronic J. {\bf 1} (1978)
155.}\ref\revHig{For a review see J.F. Gunion, H.E. Haber, G. Kane and S.
Dawson, ``{\it The Higgs Hunter's Guide}," (Addison-Wesley, 1990).}.

Another indication is that the current observed top quark mass bound
($m_t > 91$GeV)\PDG\ has already exceeded the upper bound required by the
Coleman-Weinberg mechanism of the single Higgs case to radiatively induce
the electroweak symmetry breaking\ref\rCW{S. Coleman and E. Weinberg,
Phys. Rev. {\bf D7}(1973)1888.},
which requires $m_t < 78$GeV\revHig.
Although it is rather
sufficient to have tree level breaking for particle physics purpose, it will be
difficult to have cosmological implications without radiative breaking.
However in the two-doublet case this upper bound can become sufficiently
high due to the contribution of other scalar field masses
\ref\Sher{M. Sher, Phys. Rep. {\bf 179} (1989) 273}, so that we can avoid such
a trouble.

In exchange of these positive points, having one more Higgs doublet will
introduce further  complication to the theory. Needless to say, first,
we have to deal with more observable massive scalar particles.
Theoretically, it also introduces more free parameters.
To spontaneously break the $SU(2)\times U(1)_Y$ symmetry
down to the $U(1)_{em}$ each Higgs
gets its own vacuum expectation value (\VEV), say $v_1, v_2$.
These \VEV s are phenomenologically important
but unfortunately they are not determined theoretically except in some no-scale
models\ref\rNano{A.B. Lahanas and D.V. Nanopoulos, Phys. Rep. {\bf 145} (1987)
1; Also see, for example, J.~Lopez, D.V.~Nanopoulos and A.~Zichichi,
Texas A\&M preprint, CTP-TAMU-68/92 (1992).}.
The geometric sum $v^2/2=v_1^2+v_2^2$ can be determined in terms of the
mass of the gauge boson, where $v$ denotes the electroweak symmetry breaking
scale. This however leaves the ratio of the two \VEV s,
$\tan\beta\equiv v_2/v_1$,  still undetermined.

If the two-doublet model would turn out to explain the electroweak symmetry
breaking, eventually future experiments will determine $\tan\beta$. That will
however still leave us a question why that ratio should be different from
others. This is fairly a common situation.
We always wonder if nature selects out some particular property among many
possible choices. Then curiosity drives us to look for some explanation.
Pursuing such a question, we are often led to a new phenomenon in physics.
Thus it is very important to look for any argument to constrain the
ratio rather theoretically, if possible.

With such a motivation in mind, in this paper we shall attempt to find any
relation to constrain  $\tan\beta$ in two-Higgs systems.
The result is indeed positive
and we find that there is a simple formula to express $\tan\beta$ in terms of
the couplings of the Higgs potential, so far as nature admits certain
vacuum defects during the electroweak phase transition.
Preliminary results were presented in%
\ref\Mymhig{H.S. La, Texas A\&M preprints, CTP-TAMU-73/92 (hep-ph/9211215)
(1992); CTP-TAMU-84/92 (hep-ph/9212286) (1992).}.

To demonstrate how it works we shall first work on a simple
(1+2)-dimensional toy model with $U(1)$ gauge symmetry.
Vortex solution in this model takes the role of a necessary vacuum defect.
The generic structure
however persists in the two-Higgs-doublet standard model, in which
(cosmic) string-like solution takes the role..

The interest in the application of vortex was initiated in the study of the
Ginzburg-Landau theory of the superconductivity%
\ref\Abri{A.A. Abrikosov, Sov. Phys. JETP {\bf 5} (1957) 1174.}.
Subsequently, in the relativistic field theory context, known as the
Abelian-Higgs model, the vortex lines are interpreted as string-like objects%
\ref\NieOl{H.B. Nielsen and P. Olesen, Nucl. Phys. {\bf B61} (1973) 45.}.
Due to the difficulty of solving the nonlinear field equations
any exact vortex solutions are not yet known  in a general case.
However, asymptotic solutions can be easily found. It is also known that if the
gauge coupling constant and the coupling in the Higgs potential satisfy a
special relation, one can obtain an exact solution%
\ref\VeSc{H.J. de Vega and F.A. Schaposnik, Phys. Rev. {\bf D14} (1976) 1100.}.
In the Abelian-Higgs model one can naively expect that these vortices are
stable because of the nontrivial topological configuration, {\it i.e.}
$\pi_1(U(1))\neq 0$. But more careful
analysis tells us that this is true only if the above two coupling constants
satisfy a certain inequality%
\ref\BoVe{E.B. Bogomol'nyi, Sov. J. Nucl. Phys. {\bf 24} (1976) 449.}.
If not, solutions in the higher winding sector than one are unstable.

Such a structure persists even in more realistic models in particle physics.
It was pointed out that the standard model also admits such a vortex solution
(more precisely, string-like solution which forms a vortex on a plane
perpendicular to the string.)%
\ref\Namb{Y. Nambu, Nucl. Phys. {\bf B130} (1977) 505.}%
\ref\HuTi{K. Huang and R. Tipton, Phys. Rev. {\bf D23} (1981) 3050.}%
\ref\Vach{T. Vachaspati, Phys. Rev. Lett. {\bf 68} (1992) 1977;
T.~Vachaspati and M.~Barriola, Phys. Rev. Lett. {\bf 69} (1992) 1867.}.

In this case, there is no obvious topological configuration because the
hypercharge $U(1)_Y$ is not spontaneously broken, but rather
$SU(2)\times U(1)_Y$ is broken to $U(1)_{em}$.
In ref.\Namb, however, a solution of
a pair of $SU(2)$ magnetic monopoles connected by a string is derived.
(a similar solution can also be obtained in our case. See the section 4.)
It was also pointed out that in the standard model case these monopoles can
stabilize this system. One can also derive string-like solutions in the
standard model with one Higgs doublet without
attaching to monopoles\Vach. These are unstable solutions except for
$\sin\theta_W=1$, where $\theta_W$ is the Weinberg angle.\foot{It has been
recently argued that an extra scalar field can stabilize the Z-string if it
forms a bound state\ref\VaWa{T. Vachaspati and R. Watkins, Tufts preprint,
TUTP-92-10 (1992); T. Vachaspati, TUTP-92-12 (1992).}.}.

In this paper we shall extend the results to the cases with more Higgs scalars.
If there are two Higgs multiplets to start with, we call it a two-Higgs system.
As soon as we introduce more scalar fields, the potential to induce spontaneous
symmetry breaking becomes more complicated. In particular there are more
couplings including the interactions between different scalar fields.
The structure of the phase transition itself may become more involved.
For example, in a simpler case without gauge couplings and
only two scalar couplings there are already three different critical points%
\ref\Amit{D.J. Amit, ``{\it Field Theory, Renormalization Group, and Critical
Phenomena}," (2nd ed.) (World Scientific, 1984).}.
If the two VEVs are very much different, each scalar can get its VEV one by
one\foot{The $v_2\gg v_1$
case was studied in  ref.\ref\GMM{H. Georgi, A. Manohar and G. Moore, Phys.
Lett. {\bf B 149} (1984) 234.}\ by integrating out the heavy scalar modes.}.
In the two-doublet model case, if so,
we can naively anticipate that the cosmic phase transition
would occur in two steps.
Thus perhaps a full investigation of the structure of the
fixed points of two-Higgs potential may be necessary.

This paper is organized as follows:
In sect.2 $(1+2)$-dimensional $U(1)$ gauge
models are considered. One with $U(1)\times U(1)$ global symmetry and another
without any larger accidental symmetry. Then it is shown that these vortices
are topologically stable and in some case the Bogomol'nyi bound is saturated.
In sect.3 the two-Higgs-doublet standard model is investigated.
In sect.4
it is also speculated the possibility of stabilization of the electroweak
Z-string by attaching monopoles.
In sect.5 the essence of this paper is summarized and further issues are
discussed. Finally, in the appendix A we show that the assumption
$c_1/c_2=v_1/v_2$ we made in previous sections is reasonable.

\newsec{(1+2)-d Models}

As simplified models we shall first
consider local $U(1)$ gauge theories with two Higgs singlet scalars.
In some sense these models can be viewed as a generalization of the
Abelian Higgs model in the Euclidean two-dimension, but for our purpose which
requires spontaneous symmetry breaking, we consider them rather as
$(1+2)$-dimensional systems. However solving classical field equations are
completely equivalent because we are interested in static solutions.

The Higgs potential we use in fact is motivated by
the general two-Higgs potential that
induces $SU(2)\times U(1)_Y\to U(1)_{em}$ symmetry breaking, which
we shall use in the next section and is usually
written in terms of two Higgs doublets\GerH\revHig.
In terms of singlet Higgs scalars the potential takes the following form:
\eqn\ehigpot{\eqalign{
V(\phi_1,\phi_2)=&{\la_1\over 4}(|\phi_1|^2-v_1^2)^2+
{\la_2\over 4}(|\phi_2|^2-v_2^2)^2+{\la_3\over
4}(|\phi_1|^2+|\phi_2|^2-v^2)^2\cr
&+{\la_5\over 2}\left|\phi_1^\dagger\phi_2-v_1 v_2\right|^2,\cr}}
where $v^2=v_1^2+v_2^2$. The significance of the $\la_5$-term resides in the
symmetry structure of the Higgs potential. Thus we consider the two cases for
$\la_5=0$ and $\la_5\neq 0$ separately.

Note that if $\la_1=\la_2=0$, this Higgs system has an accidental $SU(2)$
global symmetry if we require that $(\phi_1, \phi_2)$ form an $SU(2)$
doublet. This pattern of symmetry breaking, $SU(2)_{{\rm global}} \times
U(1)_{{\rm local}} \to U(1)_{{\rm global}}$, is known to lead to semilocal
topological defects\ref\VaAc{T. Vachaspati and A. Ach\'ucarro, Phys. Rev.
{\bf D44} (1991) 3067; T. Vachaspati, Phys. Rev. Lett. {\bf 68} (1992) 1977.}.
If $\la_3=0$, this becomes simply a decoupled two-scalar system with global
$U(1) \times U(1)$ symmetry. There is a vortex solution trivially
generalized from the result in ref.\NieOl.

In this paper we shall stick to the general case that $\la_3\neq 0$ and
at least one of $\la_1$ or $\la_2$ is not zero. Then we shall find that this
system reveals a rather interesting result, which cannot be obtained
otherwise.
The key observation is that the spontaneous symmetry breaking of Eq.\ehigpot\
leads to a vortex solution, whose existence will introduce an extra condition
on the Higgs \VEV s.
Then we can determine them completely, which are related only by $v^2 =
v_1^2 + v_2^2$ otherwise.

\bigskip
\leftline{{\bf Case I: $\la_5=0$}}

This two-Higgs system has not only the local $U(1)$ symmetry but
also (accidental) global $U(1)_1\times U(1)_2$ symmetry respectively for
$\phi_1$ and $\phi_2$.\foot{Such a phenomenon that the potential has a
larger global symmetry than the gauge symmetry was called accidental
symmetry first by Weinberg in
\ref\Wein{S. Weinberg, Phys. Rev. Lett. {\bf 29} (1972)
1698; Also see S. Coleman, ``{\it Aspects of Symmetry}" (Cambridge, 1985).}}
If $\la_5$ did not vanish, there would not be such a global symmetry.
In this case this model can also be seen from a different viewpoint.
One can start from a two-scalar field theory with global $U(1)\times U(1)$
symmetry with a given scalar potential. Then gauge a $U(1)$ subgroup of the
global symmetry. Such a viewpoint is common in dynamical symmetry breaking
models.

When the two scalar fields get VEVs, not only
the local $U(1)$ symmetry is spontaneously broken but also the global
$U(1)\times U(1)$ symmetry is broken. There are two Goldstone bosons due to the
global symmetry breaking. One will be absorbed to make the gauge boson massive
by the Higgs mechanism of the local $U(1)$ breaking. But, as a result,
there is one Goldstone boson left over.
Since the $\la_5$-term explicitly breaks the global symmetry, we can expect
that with $\la_5$-term it will become a pseudo-Goldstone boson whose mass is
proportional to $\la_5$. This will be considered as Case II.

Let us now consider the Lagrangian density in (1+2)-dimensional space-time
\eqn\elagr{\CL=-{\textstyle{1\over 4}}F^{\mu\nu}F_{\mu\nu}
+\half|D_\mu\phi_1|^2+\half|D_\mu\phi_2|^2
	-V(\phi_1, \phi_2),}
where $F_{\mu\nu}=\pa_\mu A_\nu-\pa_\nu A_\mu$ and $D_\mu=\pa_\mu-ieA_\mu$.
For minimum energy solutions we usually solve the minimum energy condition, but
in this case it is not so easy to obtain the necessary Bogomol'nyi bound.
Thus we shall deal with the field equations, then look for static finite energy
solutions.

The equations of motion can be obtained from the Lagrangian density as
\eqna\eomi
$$\eqalignno{D^\mu D_\mu\phi_1 &+(\la_1+\la_3)(|\phi_1|^2-v_1^2)\phi_1
		+\la_3(|\phi_2|^2-v_2^2)\phi_1=0, &\eomi a\cr
	D^\mu D_\mu\phi_2 &+(\la_2+\la_3)(|\phi_2|^2-v_2^2)\phi_2
		+\la_3(|\phi_1|^2-v_1^2)\phi_2=0, &\eomi b\cr
\pa^\mu F_{\mu\nu} &= J_\nu\equiv J_{1\nu} + J_{2\nu}, &\eomi c\cr
J_{i\nu} &=-\half ie (\phi_i^*\pa_\nu\phi_i-\phi_i\pa_\nu\phi_i^*)
-e^2 A_\nu |\phi_i|^2, \ \ i=1,2. \cr}
$$

For time-independent solutions we choose $A_0=0$ gauge, then the system
effectively reduces to a two-dimensional one. In this case since
we are interested in vortex solutions in $\IR^2$,
it is convenient to represent them in the polar coordinates
$(r, \theta)$\NieOl\ such as
\eqn\eans{\phi_1=\e^{im\theta} f(r),\ \  \phi_2=\e^{in\theta} g(r),\ \
	{\vec A}={\hat e_\theta} {1\over r}A(r),}
where $m,n$ are integers identifying each winding sector.
To become  desired finite-energy
defects located at $r=0$ these should satisfy the
following boundary conditions:
\eqn\ebc{\eqalign{&f(0)=0,\ \ g(0)=0,\ \ A(0)=0,\cr
	&f\to v_1,\ g\to v_2, \ A\to {\rm const.}\ \ {\rm as}\ \
r\to\infty.\cr}}
The constant for the asymptotic value of $A$ will be determined properly later.

In the polar coordinates the equations of motion Eq.\eomi{a-c}\
can be rewritten as
\eqna\eomii
$$\eqalignno{
-{1\over r}\pa_r(r\pa_r f) +{1\over r^2}f(m-eA)^2 + (\la_1 +\la_3) (f^2-v_1^2)f
+\la_3 (g^2-v_2^2)f &=0,\ \ \ \ \ \ \  &\eomii a\cr
-{1\over r}\pa_r(r\pa_r g) +{1\over r^2}g(n-eA)^2 + (\la_2 +\la_3) (g^2-v_2^2)g
+\la_3 (f^2-v_1^2)g &=0,\ \ \ \ \ \ \   &\eomii b\cr
-\pa_r^2 A
+ {1\over r}\pa_r A-e\left[(m-eA) f^2 + (n-eA) g^2\right] &=0.\ \ \ \ \ \ \
&\eomii c\cr}
$$
In general it will be a formidable task to solve these equations exactly, but
it is  good enough to find approximate solutions for large $r$ to show the
existence of the vortex solutions.
Imposing the boundary conditions at large $r$,  Eqs.\eomii{a,b}\ become
consistent only if $m=n$ and that it fixes the asymptotic value
$ A\to {n\over e}$ as $r\to\infty$.
This implies that there is no vortex solution of different winding numbers for
different Higgs fields. This is in fact an anticipated result because the
vortex solution we are interested in is due to the spontaneous symmetry
breaking of the local $U(1)$.
With this condition of winding numbers we can solve
Eq.\eomii{c} for large $r$ to obtain
\eqn\solA{A\to {n\over e}-n\sqrt{{\pi v\over 2e}}{\sqrt r} \e^{-r/\la}+\cdots,}
where $\la=1/ev$ is the characteristic length of the gauge field.
Note that the characteristic length defines the region over which the field
becomes significantly different from the value at the location of the defect.

Now  let us determine the characteristic lengths for
$\phi_1$ and $\phi_2$ again for large $r$ as follows.
For simplicity we consider $n=1$ case, but the result does not
really depend on $n$.
Asymptotically we look for solutions of the form
\eqn\eI{f -v_1\sim c_1\e^{-r/\xi_1},\ \ g-v_2 \sim c_2\e^{-r/\xi_2},}
where the constant coefficients $c_1$ and $c_2$  are in principle calculable.
The dimensionless part of the exact values of $c_1$ and $c_1$ should be at
most related to the dimensionless parameters of the system or may be pure
numerical constants. Thus the essential part of the following argument would
not be much changed even though we had exact solutions.
Also for our purpose only the ratio is relevant.
Therefore here we shall assume these constants
as $c_1=-v_1$ and $c_2=-v_2$, which is in a good approximation based on the
numerical study in the Abelian Higgs model%
\ref\JaRe{L. Jacobs and C. Rebbi, Phys. Rev. {\bf B19} (1979) 4486.}.
Later when we derive the Bogomol'nyi bound, we shall also confirm that this is
reasonable. (Also see the appendix A.)

Then from Eq.\eomii{a,b}\ with Eq.\eI, in the leading order we obtain
\eqna\eII
$$\eqalignno{ 0&=v_1\e^{-r/\xi_1}
\left[-{1\over \xi_1^2}+2(\la_1+\la_3)v_1^2\right] +2\la_3 v_1 v_2^2
\e^{-r/\xi_2}+\cdots , &\eII a\cr
0&=v_2\e^{-r/\xi_2}
\left[-{1\over \xi_2^2}+2(\la_2+\la_3)v_2^2\right] +2\la_3 v_1^2 v_2
\e^{-r/\xi_1}+\cdots, &\eII b\cr}
$$
where the ellipses include terms which vanish more rapidly as $r\to\infty$.

If $\xi_1\neq\xi_2$, then $\la_3$ should vanish. Therefore
to have any nontrivial solution for $\la_3\neq 0$ we are forced to identify
$$\xi\equiv\xi_1=\xi_2,$$
and that we get the desired result by demanding the vanishing coefficient
of $\e^{-r/\xi}$ in Eqs.\eII{a,b}\ as
\eqn\eresult{\tan\beta\equiv {v_2\over v_1}=\sqrt{{\la_1\over
\la_2}},\ \ \  \la_3\neq 0.}
Thus we have determined the ratio of the two Higgs \VEV s in terms of the
couplings in the Higgs potential. This tells us that although different Higgs
field gets different \VEV s, their characteristic lengths should be the same to
form a single defect. Both Higgs should reach the true vacuum at the same
distance. To do that the two \VEV s should satisfy a proper relation, which is
Eq.\eresult.

Furthermore, together with $v$, we can completely determine the \VEV s as
\eqn\eIII{
v_1 =v\,\cos\beta=v{\sqrt{{\la_2\over \la_1+\la_2}}},\ \
v_2 =v\,\sin\beta=v{\sqrt{{\la_1\over \la_1+\la_2}}}.}
The characteristic lengths $\xi_1, \xi_2$ now read
\eqn\echar{\xi\equiv\xi_1=\xi_2={1\over {\sqrt{2}}v}
\sqrt{{\la_1+\la_2\over \la_1\la_2+\la_2\la_3+\la_3\la_1}}.}
Note that, although $\tan\beta$ does not depend on $\la_3$, it is crucial to
have nonvanishing $\la_3$ coupling to obtain such a result.
The gauge boson mass is $M_A=1/\la=ev$
after spontaneous symmetry breaking. In the next section when a similar
structure is applied to  a realistic model, this gauge field in
fact can be identified with a massive neutral gauge boson, e.g. $Z^0$.

In this case one can compute the two Higgs masses to obtain
\eqn\ehmas{m_1=2v^2{\la_1\la_2+\la_2\la_3+\la_3\la_1 \over \la_1+\la_2},
\quad m_2=2v^2{\la_1\la_2\over \la_1+\la_2}.}
Note that although the characteristic lengths are the same, the masses of the
mass eigenstates of the Higgs fields are different.
The difference of the two masses is $2\la_3 v^2$.

The role of the Goldstone boson in this case perhaps need more investigation,
which we shall leave for future study.

\bigskip
\leftline{{\bf Case II: $\la_5\neq 0$}}

The $\la_5$-term explicitly breaks the global $U(1)\times U(1)$ symmetry, so
that now the Higgs potential has the same symmetry as local $U(1)$
symmetry. Thus there should not be any Goldstone boson at all. In fact one can
easily see there is another massive scalar field whose mass is proportional to
$\la_5$, signalling that it becomes a Goldstone boson at $\la_5=0$ limit.
This is a pseudo-Goldstone boson.

The $\la_5$-term in Eq.\ehigpot\ only modifies Eqs.\eomi{a,b}.
Eq.\eomi{a} picks up $\la_5 (\phi_2^\dagger\phi_1 - v_1 v_2) \phi_2$
as Eq.\eomi{b} does  $\la_5 (\phi_1^\dagger\phi_2 - v_1 v_2) \phi_1$.
Thus for $\la_5\neq 0$  we obtain the asymptotic equations similar to
Eqs.\eII{a,b}\  as
\eqna\eIIfiv
$$\eqalignno{0=&v_1\e^{-r/\xi_1}
\left[-{1\over \xi_1^2}+2(\la_1+\la_3)v_1^2\right] &\cr
&+2\la_3 v_1 v_2^2
\e^{-r/\xi_2}+\la_5 v_1 v_2^2\left(\e^{-r/\xi_1}+\e^{-r/\xi_2}\right)
+\cdots ,&\eIIfiv a\cr
0=&v_2\e^{-r/\xi_2}
\left[-{1\over \xi_2^2}+2(\la_2+\la_3)v_2^2\right]  &\cr
&+2\la_3 v_1^2 v_2
\e^{-r/\xi_1}+\la_5 v_1^2 v_2\left(\e^{-r/\xi_1}+\e^{-r/\xi_2}\right)
+\cdots. &\eIIfiv b\cr}
$$

Since $2\la_3+\la_5\neq 0$ (recall that we assumed all the couplings are
nonnegative to obtain the minimum at $v_1$ and $v_2$.), again we have to
identify the two characteristic lengths $\xi\equiv\xi_1=\xi_2$
to have nontrivial solutions. Thus the generic
structure does not really depend on the detail of the Higgs potential so far
as $\la_3>0$. However, there is one more condition on $\la_5$ we have to
impose. From Eqs.\eIIfiv{a,b} for $\xi_1 =\xi_2\equiv\xi$ we obtain
\eqn\efvre{(\la_1 -\la_5) v_1^2 = (\la_2-\la_5) v_2^2.}
Both sides must have the same sign to be consistent. This implies that if
$\la_5$ is either bigger than $\la_1, \la_2$ or smaller than them, which can be
denoted by an inequality
\eqn\elaine{(\la_5-\la_1)(\la_5-\la_2)>0.}
{}From the symmetry's point of view it is more natural to assume
the latter case so that
the $\la_5$-term would break the global symmetry rather softly. This will also
lead to a light pseudo-Goldstone boson.

The above condition Eq.\efvre\
is again nothing but the constraint on the ratio of the two
VEVs such that
\eqn\efivres{\tan\beta=\sqrt{\la_1-\la_5\over \la_2-\la_5}.}
Again it does not depend on $\la_3$ as before.
We can also compute the VEVs and the Higgs mass
in this case, but the results are equivalent to the doublet case, which will be
discussed in the next section.

\bigskip
\leftline{{\bf Remarks on the stability}}

Let us first check if the solutions we obtained in this section are
topological. This can be
understood by investigating the topology of the vacuum manifold.
In the Abelian Higgs model, where there is no accidental symmetry, it is simple
to identify the vacuum manifold by just looking at the manifold of the
equivalent vacuum states of the scalar field. Then each topological sector can
be identified by measuring the $U(1)$ flux around the vortex.

In our case if $\la_5 \neq 0$  we can also adopt the same philosophy because
there is no accidental symmetry. If we represent $\phi_1 = |\phi_1|
\e^{i\theta_1}$ and $\phi_2 = |\phi_2| \e^{i\theta_2}$, then the $\la_5$-term's
vacuum condition $\phi_1^\dagger\phi_2 - v_1 v_2 =0$ forces that $\theta_1 =
\theta_2$. Thus the vacuum manifold is still topologically $S^1$.
For vortex solutions $\pi_1 (S^1) = \IZ$ so that these are topological
solutions.

If $\la_5=0$, the topology of the vacuum manifold is a torus, $S^1\times S^1$.
One may think that since $\pi_1(S^1\times S^1) = \IZ\oplus\IZ$, the
solutions in this case should be also topological. Without the local $U(1)$
symmetry this could be true. However, due to the local symmetry the detail is
different. Recall that Eqs.\eomii{a,b}\ are consistent only if the two winding
numbers are the same. Thus the topology relevant to the vortex solutions is in
fact the diagonal $S^1$ of $S^1 \times S^1$. Although the solutions are
topological, but the topology is not completely dictated by the Higgs
potential. They are dictated by the topology of the local symmetry,
$\pi_1(U(1)) = \IZ$. This is why the two winding numbers must be the same.

Anyhow, since these solutions are topological, they are topologically stable.
However as usual it does not seem to be easy to derive the Bogomol'nyi bound
for the arbitrary couplings. However if the couplings satisfy some relation,
we can saturate the Bogomol'nyi bound.

This can be understood more easily if we use the condition $\xi_1=\xi_2$, which
can be rewritten as
$${\phi\over v}\equiv{\phi_1\over v_1}={\phi_2\over v_2}.$$
Although we have only obtained this condition asymptotically, we shall see this
indeed leads to vortex solution. We shall also find out that it is a necessary
condition to have monopole solution in sect.4. Thus we believe this will
survive as a condition for exact vortex solutions.

We will only check the Case I  explicitly, but in all the other cases it can be
done straightforwardly. With the above condition as we can easily see,
the field equations Eqs.\eomi{a-b}\ reduce to the field equations of
a single Higgs with the VEV $v=\sqrt{v_1^2+v_2^2}$ and the Higgs potential
$$V(\phi)={\textstyle {1\over 4}}\la (|\phi|^2-v^2),$$
where
$$\la\equiv {\la_1\la_2+\la_2\la_3+\la_3\la_1\over \la_1+\la_2}.$$

In the single Higgs case, as pointed out in \BoVe, if $\beta_B=2\la/e^2=1$, it
saturates the Bogomol'nyi bound. Thus we have the critical point of the
two-Higgs case as
\eqn\etbob{\beta_B={2\over e^2}
{\la_1\la_2+\la_2\la_3+\la_3\la_1\over \la_1+\la_2}=1.}
Thus these are stable solutions except for $n\geq 2$ if $\beta_B>1$.

\bigskip

This completes the proof of the existence of vortex solutions in the two-Higgs
system with the potential Eq.\ehigpot. The consistency condition of such
existence led us to be able to determine the ratio of the two Higgs \VEV s,
which otherwise are not completely determined. Much work is needed to find
exact vortex solutions, but at this moment we still get nontrivial physical
implication with approximate solutions.

Note that such a vortex solution in the $(1+2)$-dimensional space is nothing
but
the cylindrically symmetric
string-like solution in the $(1+3)$-dimensional space-time.
Thus we can expect that similar structure should exist in $(1+3)$-dimensional
models. This will be investigated in the next section.

\newsec{Two-Higgs-Doublet Standard Model}

One of the most mysterious parts of the electroweak theory lies in the Higgs
sector. Higgs was introduced to achieve the electroweak symmetry breaking
without spoiling the consistency of the theory.
In this section we shall investigate the structure of a string-like defect
(so-called ``Z-string"), which can be formed during the electroweak phase
transition, in the two-Higgs-doublet standard model, expecting
a similar structure will appear as in the $U(1)$ case considered in the
previous section.

Note that the topology of the local $SU(2)\times U(1)/U(1)$ is the same as that
of $SU(2)$, so $\pi_1(SU(2)\times U(1)/U(1)) = 0$. As is well known,
the string-like solitonic solutions we get would not be topological. This is
different from the $U(1)$ case. However it is known that there are
nontopological solitons in the standard model\Namb\HuTi\Vach.

We shall use the CP invariant two-doublet Higgs potential that
induces $SU(2)\times U(1)_Y\to U(1)_{em}$ symmetry breaking\GerH\revHig :
\eqn\edhigpot{\eqalign{
V(\phi_1,\phi_2)=&\lfr{1}{2}\la_1\left(|\phi_1|^2-v_1^2\right)^2+
\lfr{1}{2}\la_2\left(|\phi_2|^2-v_2^2\right)^2+
\lfr{1}{2}\la_3\left(|\phi_1|^2+|\phi_2|^2-v_1^2-v_2^2\right)^2 \cr
&+\la_4\left(|\phi_1|^2 |\phi_2|^2-|\phi_1^\dag\phi_2|^2\right)
+\la_5\left|\phi_1^\dag\phi_2 -v_1 v_2\right|^2,\cr}}
where $\phi_1,\ \phi_2$ are $SU(2)$ doublets.
In this section we shall stick to the general case that $\la_i\neq 0$
for $i=1,2,3$ and also assume that all $\la_j,\ j=1,\ldots, 5$ are nonnegative.
This potential shows $\phi_1, \phi_2\leftrightarrow -\phi_1, -\phi_2$
discrete symmetry, which is
necessary to suppress the flavor changing neutral current.
Then we shall find that this system reveals a rather interesting result,
which cannot be obtained otherwise.

Note that if $\la_4=0=\la_5$, the Higgs potential has a global $U(2)\times
U(2)$ symmetry. The symmetry breaking will lead to global
$U(1)\times U(1)$ unbroken
so that there will be two Goldstone bosons left over. If only $\la_5$ is
vanishing, due to $|\phi_1^\dagger\phi_2|^2$ in the $\la_4$-term
the global symmetry now becomes
$U(2)\times U(1) \times U(1)$. There still is one Goldstone boson after
symmetry breaking to global $U(1)\times U(1)$.

There however are good reasons to keep $\la_5\neq 0$.
First, we don't want Goldstone bosons which may not be welcomed
phenomenologically. Secondly, though not directly related,
in the supersymmetric models $\la_5$ is related to the supersymmetry breaking
parameter so that $\la_5\neq 0$ to have the supersymmetry broken.
Note that for $\la_5\neq 0$ there is a global $U(2)$ symmetry, which is locally
isomorphic to $SU(2)\times U(1)$.

Let us consider the bosonic sector of the standard model
described by the Lagrangian density
\eqn\edlagr{
\CL=-\half\tr G_{\mu\nu}G^{\mu\nu}-{\textstyle{1\over 4}}F^{\mu\nu}F_{\mu\nu}
+|D_\mu\phi_1|^2+|D_\mu\phi_2|^2
	-V(\phi_1, \phi_2),}
where $F_{\mu\nu}=\pa_\mu B_\nu-\pa_\nu B_\mu$,
$G_{\mu\nu}^a=\pa_\mu W_\nu^a-\pa_\nu W_\mu^a + g\eps^{abc} W_\mu^b W_\nu^c$,
and $D_\mu=\pa_\mu-ig'{Y\over 2}B_\mu-ig{\tau^a\over 2} W_\mu^a$.
Both Higgs' have hypercharge $Y=1$.

Then the equations of motion for the scalar fields are
\eqna\edomi
$$\eqalignno{
0=&D^\mu D_\mu\phi_1 +\la_1\left(|\phi_1|^2-v_1^2\right)\phi_1
	+\la_3\left(|\phi_1|^2+|\phi_2|^2-v_1^2-v_2^2\right)\phi_1 &\cr
&+\la_4\left(|\phi_2|^2\phi_1-(\phi_2^\dag\phi_1)\phi_2\right)
	+\la_5(\phi_2^\dag\phi_1-v_1 v_2)\phi_2,&\edomi a\cr
0=&D^\mu D_\mu\phi_2 +\la_2\left(|\phi_2|^2-v_2^2\right)\phi_2
	+\la_3\left(|\phi_1|^2+|\phi_2|^2-v_1^2-v_2^2\right)\phi_2 &\cr
&+\la_4\left(|\phi_1|^2\phi_2-(\phi_1^\dag\phi_2)\phi_1\right)
	+\la_5(\phi_1^\dag\phi_2-v_1 v_2)\phi_1, &\edomi b\cr}
$$
and for the gauge fields we have
\eqna\edomic
$$\eqalignno{
&-\pa^\mu F_{\mu\nu}=j_\nu\equiv j_{1\nu} + j_{2\nu}, &\edomic a\cr
&\quad j_{i\nu} \equiv\lfr{1}{2}ig' \left[\phi_i^\dag\pa_\nu\phi_i
-(\pa_\nu \phi_i)^\dag\phi_i\right]
+\lfr{1}{2}g'^2 B_\nu |\phi_i|^2+\lfr{1}{2}g'g W_\nu^a\phi_i^\dag\tau^a\phi_i,
\ \ i=1,2,&\cr
&-\pa^\mu G^a_{\mu\nu} -g\eps^{abc}W^{b\mu}G^c_{\mu\nu}
=J^a_\nu\equiv J^a_{1\nu} + J^a_{2\nu}, &\edomic b\cr
&\quad J^a_{i\nu} \equiv\lfr{1}{2} ig \left[\phi_i^\dag\tau^a\pa_\nu\phi_i
-(\pa_\nu \phi_i)^\dag\tau^a\phi_i\right]
+\lfr{1}{2}gg' B_\nu \phi_i^\dag\tau^a\phi_i
+\lfr{1}{2}g^2W_\nu^b\phi_i^\dag\tau^a\tau^b\phi_i,
\ \ i=1,2, \cr}
$$
For time-independent solutions we choose $B_0=0=W_0^a$ gauge and impose the
cylindrical symmetry around the string, then the system
effectively reduces to a two-dimensional one. In this case the string solutions
in the $(1+3)$-dimensional spacetime correspond to
the vortex solutions in $\IR^2$.
When Higgs gets \VEV, the false vacuum region forms vacuum defects.
As usual,  we redefine the neutral gauge fields as
\eqn\azfld{
A_\mu=\cos\theta_W B_\mu+\sin\theta_W W_\mu^3,\quad
Z_\mu=\sin\theta_W B_\mu-\cos\theta_W W_\mu^3,}
where $\theta_W$ is the Weinberg angle defined by $\tan\theta_W=g'/g$.
We shall also use $\gt\equiv \half\sqrt{g^2+g'^2}$ for convenience.

For vortex solutions
it is convenient to represent them in the polar coordinates
$(r, \theta)$\NieOl\ such as
\eqn\edans{\phi_1=\pmatrix{0\cr\e^{im\theta} f_1(r)\cr},
\quad \  \phi_2=\pmatrix{0\cr \e^{in\theta} f_2(r)\cr},\quad \
	{\vec Z}={\hat e_\theta} {1\over r}Z(r),}
where $m,n$ are integers identifying each ``winding'' sector ( we shall
come back to this point later again.).
Here we are mainly interested in the case of $W_\mu^1=0=W_\mu^2$,
but we expect there are other solutions
similar to the case of ref.\Vach.

Then Eqs.\edomic{a,b}\ become
\eqna\edmiia
$$\eqalignno{
0=&-{1\over r}\pa_r(r\pa_r B_\theta)+{1\over r^2}B_\theta &\cr
&-{g'\over r}\left[\left(m-\half(g'B_\theta-g W_\theta^3)\rbr f_1^2
+ (n-\half(g'B_\theta-g W_\theta^3) )f_2^2\right] , &\edmiia a\cr
0=&-{1\over r}\pa_r(r\pa_r W^3_\theta)+{1\over r^2}W^3_\theta &\cr
&+{g\over r}\left[\left(m-\half(g'B_\theta-g W_\theta^3)\rbr f_1^2
+ (n-\half(g'B_\theta-g W_\theta^3) )f_2^2\right] . &\edmiia b\cr}
$$
As we can easily see, $A_\mu$ satisfies a trivial equation so that we can
set $A_\mu=0$. Thus from the rest of the equations of motion we obtain
\eqna\edomiia
$$\eqalignno{
0&=-{1\over r}\pa_r(r\pa_r f_1) +{1\over r^2}f_1(m-\gt Z)^2 &\cr
&+ (\la_1 +\la_3) (f_1^2 - v_1^2)f_1
+\la_3 (f_2^2- v_2^2)f_1 +
\la_5\left(f_1f_2- v_1v_2\e^{i(n-m)\theta}\right)\! f_2,
\quad\quad &\edomiia a\cr
0&=-{1\over r}\pa_r(r\pa_r f_2) +{1\over r^2}f_2(n-\gt Z)^2 &\cr
&+ (\la_2 +\la_3) (f_2^2- v_2^2)f_2 +\la_3 (f_1^2- v_1^2)f_2 +
\la_5\left(f_1f_2- v_1v_2\e^{i(m-n)\theta}\right)\! f_1,
\quad\quad &\edomiia b\cr
0&=-\pa_r^2 Z+{1\over r}Z
-2\gt\left[(m-\gt Z) f_1^2 + (n-\gt Z)f_2^2\right]. &\edomiia c\cr}
$$
Note that $\la_4$ coupling does not take part in this structure classically.

To become  desired finite-energy defects located at $r=0$
the solutions we are looking for should satisfy the
following boundary conditions:
\eqn\ebc{\eqalign{
f_1(0)=0,\ \ f_2(0)=0,\ \ &Z(0)=0,\cr
f_1\to v_1,\ f_2\to v_2, \ &Z\to {\rm const.}\ \ {\rm as}\ \
r\to\infty.\cr}}
The constant for the asymptotic value of $Z$ will be determined properly later.

Again we shall look for asymptotic solutions.
Imposing the boundary conditions at large $r$,
Eqs.\edomiia{a,b}\
become consistent only if $m=n$ and that it fixes the asymptotic value
$ Z\to {n/ \gt}$ as $r\to\infty$.
This implies that there is no vortex solution of different ``winding''
numbers for different Higgs fields.
With this condition of winding numbers we can solve
Eq.\edomiia{c}\ for large $r$ to obtain\NieOl\
\eqn\dsolA{
Z\to {n\over \gt}-n\sqrt{{\pi v\over 2\gt}}{\sqrt r} \e^{-r/\la}+\cdots,}
where $\la=1/\gt v$ is the characteristic length of the gauge field.
Note that the characteristic length defines the region over which the field
becomes significantly different from the value at the location of the defect.

The asymptotic solutions for $\phi_1$ and $\phi_2$  can be found as follows:
For simplicity we consider $n=1$ case, but the result does not
really depend on $n$. Besides, since these are nontopological, it is not really
necessary to consider other $n$ sector.
Asymptotically we look for solutions of the form
\eqn\edI{
f_1 -v_1\sim c_1\e^{-r/\xi_1},\ \ f_2-v_2 \sim c_2\e^{-r/\xi_2},}
where $\la_1$ and $\la_2$ are the characteristic lengths of
$\phi_1$ and $\phi_2$ respectively and
the constant coefficients $c_1$ and $c_2$  are in principle calculable, thus
they are not free parameters.
Note that  we can normalize any dimensionless constants in $c_i$ to be
the same. Furthermore, for our purpose only the ratio is relevant.
Therefore these constants
can be taken as $c_1=-v_1$ and $c_2=-v_2$ in a good approximation.
Even though the exact results differed from these, the essential argument of
the
following is much the same.
Then in the leading order we obtain
\eqna\eIIa
$$\eqalignno{
v_1\e^{-r/\xi_1}
\left[-{1\over \xi_1^2}+2(\la_1+\la_3)v_1^2+\la_5 v_2^2
\right] +(2\la_3+\la_5) v_1 v_2^2
\e^{-r/\xi_2}+\cdots &=0,\quad \quad \ \  &\eIIa a\cr
v_2\e^{-r/\xi_2}
\left[-{1\over \xi_2^2}+2(\la_2+\la_3)v_2^2+\la_5 v_1^2
\right] +(2\la_3+\la_5) v_1^2 v_2
\e^{-r/\xi_1}+\cdots &=0,\ \ \ \ &\eIIa b\cr}
$$
where the ellipses include terms which vanish more rapidly as $r\to\infty$.

Recall that $\la_3>0$ and $\la_5\geq 0$ so that $2\la_3+\la_5\neq 0$.
Thus to have any vortex solution we are forced to identify the two
characteristic lengths of the scalar fields such that
$$\xi\equiv\xi_1=\xi_2.$$
We shall find in the next section that to attach monopoles at each side of this
string we should require that $\phi_1/v_1=\phi_2/v_2$. Thus this is also a
necessary condition to stabilize by attaching monopoles.

To be consistent, as in the $U(1)$ case we also obtain an inequality
\eqn\edfvr{(\la_5 -\la_1)(\la_5-\la_2)>0.}
However in the two-doublet stabdard model case $\la_5$ does not need to be
small to be natural.
Then we get the desired result by demanding the vanishing coefficients
of $\e^{-r/\xi}$ in Eqs.\eIIa{a,b}\ as
\eqn\edresult{\tan\beta\equiv {v_2\over v_1}=\sqrt{{\la_1-\la_5\over
\la_2-\la_5}},\ \ \  \la_3\neq 0\ \ {\rm or} \ \la_5\neq 0.}
Thus we have determined the ratio of the two Higgs \VEV s in terms of the
couplings in the Higgs potential. This tells us that although different Higgs
field gets different \VEV s, their characteristic lengths should be the same to
form a single defect. Both Higgs should reach the true vacuum at the same
distance. To do that the two \VEV s should satisfy a proper relation, which is
Eq.\edresult.

Furthermore, together with $v$, we can completely determine the \VEV s as
\eqn\edIII{v_1 ={v\over\sqrt{2}}
\,\cos\beta={v\over\sqrt{2}}{\sqrt{{\la_2-\la_5\over \la_1+\la_2-2\la_5}}},\ \
v_2 ={v\over\sqrt{2}}
\,\sin\beta={v\over\sqrt{2}}{\sqrt{{\la_1-\la_5\over \la_1+\la_2-2\la_5}}}.}
The characteristic lengths $\xi_1, \xi_2$, now satisfy
\eqn\edchar{\xi\equiv\xi_1=\xi_2={1\over v}
\sqrt{{\la_1+\la_2-2\la_5\over \la_1\la_2+\la_2\la_3+\la_3\la_1
-\la_5(2\la_3+\la_5)}}.}
Note that, although $\tan\beta$ does not depend on $\la_3$, it is crucial to
have nonvanishing $\la_3$ or $\la_5$ coupling to obtain such a result.
The gauge boson mass is $M_Z=1/\la=\gt v$
after spontaneous symmetry breaking.

In this two-Higgs-doublet model there are five physical Higgs bosons:
$H^{\pm},\ A^0,\ H^0,\ h^0$. $A^0$ is a CP-odd neutral scalar, while
$H^0,\ h^0$ are CP-even scalars. $h^0$ denotes the lightest Higgs.
Using Eq.\edIII, we can compute the masses of all these physical Higgs
bosons in terms of the couplings in the Higgs potential and $v$, where
$v=247$ GeV. $\la_5$ is related to $M_{A^0}$ and $M_{H_0,h_0}$ can be
determined
in terms of $\la_1,\la_2,\la_3,\la_5$ and $v$. Thus we only have five free
parameters, if such an electroweak Z-string exists.

In particular the neutral Higgs masses become impressively simple:
\eqn\edhms{\eqalign{
m_{H^0} &=v^2\left[\la_3 + {\la_1\la_2-\la_5^2\over \la_1+\la_2
-2\la_5}\right],\cr
m_{h^0} &=v^2 {2\la_1\la_2-\la_5(\la_1+\la_2)\over 2(\la_1+\la_2
-2\la_5)}.\cr}}
Note that $m_{h^0}$ does not depend on $\la_3$.

The appearance of integers in the solutions,
which we still call ``winding'' number,
is rather intriguing because there is no
explicit $U(1)$ symmetry to be broken
which should determine the necessary topological sector.
If our vortex solutions are nontopological as in ref.%
\ref\tdlee{T.~D.~Lee, Phys. Rep. {\bf 23} (1976) 254.},
there should not be such a parameter.
This however
can be explained as follows: If we regard $W_\mu^1=0=W_\mu^2$ as gauge fixing
conditions, then effectively we can view
the symmetry of the system as $U(1)\times U(1)_Y$. When we twist this symmetry
to obtain $U(1)_{em}$, the remaining twisted $U(1)_{\gt}$
is spontaneously broken to lead to the winding sector.

Since $SU(2)$ is a simple group, $U(1)\times U(1)_Y$ is not an invariant
subgroup of $SU(2)\times U(1)_Y$.
Furthermore, $\pi_1\left(SU(2)\times U(1)_Y/U(1)_{em}\right) =0$
implies that these winding sectors would not provide any topological stability.
In other words, they must be gauge equivalent to $n=1$ solution via deformable
gauge transformation.

Even for $n=1$ solution it is most likely that this solution would not saturate
the Bogomol'nyi bound.
Although it obviously is a finite energy solution, it
does not seem to be a classically stable solution. It is argued that there is a
case of quantum stabilization of a classically unstable solution%
\ref\preskill{J. Preskill,  Caltech preprint, CALT-68-1787 (hep-ph/9206216).}.
It however cannot be applied in this case unless one of $\la_4$ or $\la_5$
vanishes because there is no tree level Goldstone boson.
If $\la_5=0$, then there is one Goldstone boson. But the argument still does
not apply because $\pi_1(G_{{\rm gauge}}/H_{{\rm gauge}})$ is still trivial.

Nevertheless, there is one more possibility to stabilize such an electroweak
Z-string, which will be presented in the next section.

\newsec{Monopole-String-Antimonopole}

In the one-doublet standard model it is claimed that such a string solution can
be stabilized by attaching monopoles at each end and keeping them sufficiently
far apart\Namb. The rationale behind the stabilization is that these monopoles
are genuine $SU(2)$ monopoles so that they can be topologically stable.
In other words, the string can be constructed as a string connecting two
Wu-Yang type monopoles.

In fact we can apply the same argument here. As we shall see soon, the
two-doublet model also admits such a monopole solution. The consistency
condition for the existence is again $\phi_1/v_1 = \phi_2/v_2$, which is an
equivalent condition to $\xi_1 = \xi_2$ in the string case.
Thus it seems to us that this condition is not just accidental but may have
more profound significance in the spontaneous symmetry breaking of the
two-Higgs systems.

In the static case
the energy density  of the system can be easily derived from the Lagrangian
density Eq.\edlagr. Then the minimum energy conditions are
\eqn\emnmo{D_\mu\phi_1=0,\ D_\mu\phi_2=0,}
and
\eqn\evmn{V(\phi_1,\phi_2)=0.}
Eq.\evmn\ implies that
\eqn\ppmcn{|\phi_1|^2=v_1^2,\ |\phi_2|^2=v_2^2,\ \phi_1^\dagger\phi_2=v_1 v_2,}
where the third condition does not follow from the two other conditions.

{}From Eq.\emnmo\ by multiplying $\phi_i^\dagger\tau^a$ and subtracting
its hermitean conjugate we obtain
\eqn\enbsol{
g v_i^2 W_\mu^a+ g'B_\mu (\phi_i^\dagger\tau^a\phi_i)
=-i\left(\phi_i^\dagger\tau^a
\pa_\mu\phi_i- \pa_\mu\phi_i^\dagger\tau^a\phi_i\right),\ \ i=1,2.}
These two simultaneous equations need to be consistent to have solutions.
Comparing them, we are required to identify
\eqn\ecocn{{\phi_1\over v_1} ={ \phi_2\over v_2}.}
Thus the two equations become identical.
Then using the Nambu's method we can solve Eq.\enbsol\ to obtain
$W_\mu^a$ and $B_\mu$ in terms of $\phi_1$ and $\phi_2$.

For a monopole solution with a singularity along the negative $z$-direction,
the Higgs fields would be
\eqn\mostr{
\phi_i=v_i\left({\cos\half\theta\atop\sin\half\theta \e^{i\varphi}}\right),\
\ r\neq 0,\ \ 1=1,2,}
where the singularity is identified by the ill-defined phase at $\theta=\pi$.
Thus we have
\eqn\emoph{\phi_i^\dagger\tau^a\phi_i=v_i^2{x^a\over r}.}
Since we expect that $B_\mu$ behaves like a monopole with a Dirac string,
we use
\eqn\emodis{g'B_a=-\eps^{ab3}{x^b\over r(r+z)},}
where we have identified the spatial indices and the gauge group indices as
$a,b,...=1,2,3$ and $x^1=x$, $x^2=y$, $x^3=z$. In the spherical  coordinates we
can easily see that only $\varphi$-component is nonvanishing.
$B_0=0$ by the gauge choice.

Then plugging into Eq.\enbsol\ we can solve for $W_\mu^a$ to get
\eqn\emowst{g W_a^b=-\eps^{abc}{x^c\over r^2}, }
which is nothing but the $SU(2)$ monopole solution%
\ref\tHPol{G. 'tHooft, Nucl. Phys.
{\bf B79} (1974) 276; A.M. Polyakov, JETP Lett. {\bf 20} (1974) 194.}.
Note that $\phi_i^\dagger\tau^a\phi_i$ takes the role of the adjoint
Higgs field in the usual non-Abelian Higgs model.
Since $W_a^b$ denotes a genuine $SU(2)$ monopole, the string carries
only $U(1)$ returning flux but does not carry $SU(2)$ returning flux.

Now we can patch a pair of monopole and antimonopole, which leads to
\eqn\mostpa{\phi_i=v_i\left({\cos\half\Theta\atop\sin\half\Theta
\e^{i\varphi}}\right),}
where $\cos\Theta=\cos\theta_1-\cos\theta_2 +1$ and $\theta_1$ $(\theta_2)$ is
measured from the position of the monopole (antimonopole). If the distance
between the monopole and the antimonopole is sufficiently far apart, the
solution we derived can be used for each side so that this can be viewed as
a monopole-string-antimonopole system. One can easily obtain the relevant
solutions by generalizing Eqs.\emodis\emowst\ in this case.
Thus by the same reasoning as in \Namb, this would be stable.

\newsec{Discussion}

We have shown that two-Higgs systems in general admit vortex solutions, which
requires a specific ratio of the two VEVs. This can be understood more easily
if we use the condition
$$\widehat\phi\equiv{\phi_1\over v_1}={\phi_2\over v_2}.$$
Recall that this condition not only shows up as a consistency condition of the
vortex solutions but also shows up as that of the monopole solution considered
in sect.4.

Then as we can easily see, the field equations reduce to the field equations of
a single Higgs with the VEV $v$ and the Higgs potential
$V(\phi)=\la (|\phi|^2-v^2/2)^2$, where $\phi=v\widehat{\phi}/\sqrt{2}$
(in the 2-d case replace $v^2/2$ by $v^2$) and
$$\la\equiv {\la_1\la_2+\la_2\la_3+\la_3\la_1-\la_5(2\la_3+\la_5)
\over \la_1+\la_2-2\la_5}.$$
All such results can be obtained if
$$\tan\beta\equiv {v_2\over v_1}=\sqrt{{\la_1-\la_5\over
\la_2-\la_5}},\ \ \  \la_3\neq 0.$$
These results also apply to the cases of $\la_5=0$.

Although there are characteristic differences,
in fact we have observed that so far as vortex solution is concerned,
such a structure is quite generic for both cases:
the (1+2)-dimensional $U(1)$ system and the (1+3)-dimensional $SU(2)\times
U(1)$. Two big characteristic differences are that the former is topological
and saturates Bogomol'nyi bound (in some case), whilst the latter is
nontopological but can be stabilized by attaching monopoles.

In this paper we have dealt with the tree level potential. It is usually
believed that tree level solution reappears in loop-corrected
effective potential, although there might be quantitative differences.
This is in fact relevant to study the porperty of vacuum defects in the
context of the cosmological phase transition.

For the unstable solutions we need further investigation to find out what kind
of cosmological trace they could leave. But one certain thing is that as soon
as we find out how nature selects out $\tan\beta$ (if nature prefers the
two-doublet model), we shall understand its mystery if $\tan\beta$ turns out to
meet our claim.
The ultimate proof of the existence of such vacuum defects should be determined
by experiments or by observations.

It will be also important and interesting to find out if there is any other
reason nature prefers a specific $\tan\beta$ theoretically. This will be still
a question to be answered after we determine experimentally. In this sense, we
hope this work can provide a clue to future investigation.

\bigbreak\bigskip\bigskip\centerline{{\bf Acknowledgements}}\nobreak

\par\vskip.3truein

The author would like to thank R. Arnowitt
for helpful discussions.
This work was supported in part by NSF grant PHY89-07887 and World Laboratory.

\appendix{A}{}

In this appendix we shall show explicitly why $c_1/c_2=v_1/v_2$ is a reasonable
assumption in the case I of the section 1. In other cases it can
also be generalized straightforwardly so that we leave them as simple
exercises. The  main rationale behind our assumtion is that these constants are
not usually determined by solving nonlinear differential equations
asymptotically so that one should use them as input parameters. Of course they
will be determined by solving equations exactly, however we are not yet able to
solve the equations in question exactly.
As we shall see, these will be ill-defined if we try to determine otherwise and
the assumption we made is the only reasonable consistent choice.

\def\tc{{\tilde c}}
In eq.\eI\ we can conveniently redefine $c_i=\tc_i v_i$, where $\tc_i$ are
dimensionless. Also let $\tc\equiv \tc_1/\tc_2$, then the condition for
$\xi_1=\xi_2$ reads as
\eqn\enori{\tan\beta=\sqrt{{\la_1+\la_3-\tc \la_3\over \la_2+\la_3 - {1\over
\tc}\la_3}}.}
Suppose $\tc$ is not an input parameter but to be determined by this
equation for any $\tan\beta$ to nullify our claim that this equation relates
$\tan\beta$ and other couplings, then one should be able to determine $\tc$ for
given $\tan\beta$.
In this case eq.\enori\ becomes a quadratic equation for $\tc$ as
\eqn\enorii{\la_3 \tc^2 + \left(\tan^2\beta(\la_2+\la_3) -
(\la_1+\la_3)\right)\tc -\la_3\tan^2\beta=0}
and one can obtain
\eqn\enoiii{\tc_\pm={1\over 2\la_3}\left[-\tan^2\!\beta(\la_2+\la_3)
+ (\la_1+\la_3)
\pm\sqrt{\left(\tan^2\!\beta (\la_2+\la_3)- (\la_1 +\la_3)\right)^2 +
4\la_3^2\tan^2\!\beta}\right].}

First, note that $\tc_-<0$ so that one of the Higgs fields approaches
to the true vacuum from the wrong direction. Furthermore,
$\tc_-\to 0$ for any $\la$'s as $\tan\beta\to 0$,
so $\tc_-$ leads to ill-defined coefficients.
Thus we are left with $\tc_+>0$.

Second, if $\la_3\to 0$, then any $\tc$ is ill-defined unless
the coefficient of the linear term in eq.\enorii\
vanishes. This vanishing condition is
nothing but eq.\eresult\ as $\la_3\to 0$. Otherwise, $\tc$ either increases
indefinitely or approaches to $0$ indefinitely. This is unreasonable because
for vortex solutions with the same characteristic length
we should expect $\tc_1$ and $\tc_2$ are of the similar order
for any $\la_3$. Without loss of generality, say, $\tc_1\ll\tc_2\sim 1$,
then the notion of characteristic length in the
asymptotic formula for $\phi_1$ field we assumed fails to make sense.

Therefore, eq.\enorii\ cannot be solved for plausible $\tc$ consistently for
$\tan\beta$ as an input parameter.
 This leads us to a conclusion
 that we should treat $\tc$ as an input parameter and
the assumption we make in the previous sections is reasonable.
Thus $\tc=1$, which leads to eq.\eresult\ consistently, is a
well-defined reasonable assumption to present our argument.


%
\listrefs
\vfill\eject
\bye